\documentclass{PoS}
\usepackage{amsfonts,amsbsy,amsmath,amssymb}
\usepackage{epsfig}
\usepackage{subfigure}
\title{Adjoint zero-modes as a tool to understand the Yang-Mills vacuum}

\ShortTitle{Adjoint zero-modes as a tool to understand the Yang-Mills vacuum}

\author{Margarita Garc\'{\i}a P\'erez\\
Instituto de F\'isica Te\'orica UAM-CSIC\\
Universidad Aut\'onoma de Madrid, Cantoblanco, 28049 Madrid, Spain\\
E-mail: \email{margarita.garcia@uam.es}}
\author{Antonio Gonz\'alez-Arroyo\\
Departamento de F\'isica Te\'orica and Instituto de F\'{\i}sica 
Te\'orica UAM-CSIC\\\
Universidad Aut\'onoma de Madrid, Cantoblanco, 28049 Madrid, Spain\\
E-mail: \email{antonio.gonzalez-arroyo@uam.es}}
\author{\speaker{Alfonso Sastre}\\
Departamento de F\'isica Te\'orica and Instituto de F\'{\i}sica 
Te\'orica UAM-CSIC\\
Universidad Aut\'onoma de Madrid, Cantoblanco, 28049 Madrid, Spain\\
E-mail: \email{alfonso.sastre@uam.es}}

\abstract{
The use of adjoint (quasi) zero-modes of the Dirac operator
to probe the Yangs-Mills vacuum has been recently 
advocated by Gonz\'alez-Arroyo and Kirchner.
The construction relies on the use of the super-symmetric zero mode which,
for classical configurations, provides a direct estimate of the gauge action
density. In the  lattice implementation of this idea, we show how the results
improve considerably  if the overlap operator is used instead of the
Wilson-Dirac one. Before proceeding to the detailed study of Monte Carlo
ensembles, we studied here a series of potentially complicated
situations which can be encountered. In particular, we study the case
of instanton anti-instanton pairs and analyse how the results depend
upon separation. The effect of lattice artifacts is also of concern.
Indeed, a statistical analysis of zero modes of thermalised SU(2) 
configurations at $\beta=2.57$ shows a significant fraction having
$4N+2$ adjoint zero modes, in contradiction with the index theorem. This
violation must be  associated to the roughness of the lattice
configurations. Indeed, we  show that this situation occurs
for instantons of size of the order  of the lattice spacing. }

\FullConference{The XXV International Symposium on Lattice Field Theory\\
July 30 - August 4 2007\\
Regensburg, Germany}

\begin{document}

\section{Introduction.}

In the past years a considerable effort has been devoted
to the analysis of the topological structure of pure
Yangs-Mills theory on the lattice. These studies encompass from global
quantities, like the topological susceptibility, to local
ones, in an attempt to link topology to models of chiral
symmetry breaking and confinement. For that one has to deal
with the roughness of lattice Monte-Carlo
configurations, which are generally plagued with ultraviolet
fluctuations. Several cooling and smearing algorithms have
been proposed in order to obtain a smoother image of these
configurations. Although these methods are useful in order
to compute global quantities, they are
criticized as a tool to analyse local properties. An alternative,
claimed to distort less the original ensemble,
is to use filtering methods based on the Dirac operator \cite{filter}.
The main idea is the relation between fermions and topology given by
the Atiyah-Singer index theorem and the correlation between the gauge action
density and the local density of the eigenstates of the Dirac operator.
As will be described below, this relation is particularly neat for fermions
in the adjoint representation.

In what follows we will first present a discussion of the advantages
of using filtering methods based on the adjoint Dirac operator, as
advocated in \cite{gak}. In that paper it was argued that in order to
apply the method to Monte Carlo generated configurations on the lattice
the use of the  overlap Dirac operator was desirable. Here we have
carried out this program and applied it to a set of potentially delicate
situations. The first one is that of instanton anti-instanton (IA) pairs.
The construction in \cite{gak} is exact for classical
solutions of the equations of motion and relies on the existence of
quasi zero modes of the Dirac operator. IA pairs are not of this sort, with
quasi zero modes disappearing as the separation in the pair decreases.
Still our results show how, even in the case of rather small separation, the 
topological charge density can be reconstructed in terms of the density 
of the, so called, super-symmetric modes. 
The second potentially delicate case is related to dislocations, lattice
artefacts associated to small instantons. Some time ago Edwards,
Heller and Narayanan \cite{EHN} reported on the presence of configurations
with $4N+2$ adjoint-zero modes in lattice thermalized ensembles.
This is in contrast with the continuum expectation of $4N$. We have 
also encountered a significant fraction of these cases in our ongoing 
Monte-Carlo simulations.  We will argue that they are necessarily related 
to the roughness of the lattice configurations and show it explicitly 
on a set of instantons with size of the order of the lattice spacing. 

\section{Reconstructing the action density from adjoint zero-modes.}

We will describe here the basic ideas underlying the proposal in Ref. 
\cite{gak} and what are the advantages over other filtering methods.

The construction is based on the super-symmetric zero modes
for classical solutions of the Euclidean equations of motion, which
read ~\cite{JR}:
\begin{equation}
\psi^a = \frac{1}{8}\ F_{\mu\nu}^a\ [\gamma_\mu,\gamma_\nu] \, V \, ,
\label{eq:susy}
\end{equation}
with $F_{\mu\nu}$ the gauge field strength and $V$  any constant four-spinor. From this expression two positive
and two negative chirality zero modes are obtained (note that adjoint
zero modes always come in pairs related by charge conjugation
$C$ through $\psi_c = \gamma_5 C \psi^*$).
The densities of these zero modes give respectively the self-dual
and anti-self-dual parts of the action density. Taking, for instance,
$V^+ = \left (1,0,0,0\right)$, the corresponding positive chirality  
zero mode in Eq. (\ref{eq:susy}) becomes
\begin{equation}
\psi^a_+(x) \propto \left(\begin{array}{c}E^a_3(x) + B^a_3(x) \\ E^a_1(x) + B^a_1(x) - i(E^a_2(x) + B^a_2(x))
\\ 0 \\ 0 \end{array}\right) \, ,
\end{equation}
selecting only the self-dual part of the action density. 
There is one peculiarity of $\psi^a_+$ that allows to single it out 
from the subspace of zero modes: the imaginary 
part of the first component is zero in every point $x$ and for all 
components in colour space. 

Given the fact that, for classical solutions, the super-symmetric zero modes 
trace the action density, the proposal in Ref. \cite{gak} is to 
make an analogous construction on generic configurations. 
The strategy is to look for the low lying modes of the Dirac
operator and single out the super-symmetric modes by imposing reality 
conditions as the one described above. For example, within the space of
positive chirality low modes we select the combination that minimizes
the quantity
$\sum_{x,a}[{\rm Im} \ \psi_+^{1,a}(x)]^2 $. 
Charge and action densities are reconstructed respectively
out of the sum and difference of $|\psi_+^a(x)|^2$ and $|\psi_-^a(x)|^2$.
This method is computationally cheaper than those based on the use of the
fundamental representation, since it requires a much smaller number of modes
to reconstruct the (anti) self-dual structures.

 The authors in Ref. \cite{gak} have tested the method using the
Wilson-Dirac operator. The test works well on smooth configurations.
Once moderate quantum fluctuations are included the method succeeds 
in filtering the UV noise.  Nevertheless, the Wilson-Dirac operator is
not optimal for this program since, in many cases, it does not show a clear 
gap in the spectrum.  A better option is to use the
Neuberger-Dirac operator \cite{neuberger}:
\begin{equation}
 D_{ov} = \frac{1}{2}\left(1 + \gamma_5\epsilon(H_{WD})\right)\,,
\label{eq:overlap}
\end{equation}
where  $H_{WD} = \gamma_5(D_W - m_{WD})$ and $D_W$ is the  
Wilson-Dirac operator in the adjoint representation. One advantage
of using a chiral operator is that 
we can compute each chirality independently by diagonalising the
operators:
\begin{equation}
 D\bar{D} = P_+(\gamma_5D_{ov})^2P_+= \frac{1}{2}P_+\left(1 + \epsilon \right)P_+
\end{equation}
and
\begin{equation}
 \bar{D}D = P_-(\gamma_5D_{ov})^2P_-= \frac{1}{2}P_-\left(1 - \epsilon \right)P_-\,,
\end{equation}
with $P_{\pm} = \frac{1}{2}\left(1 \pm \gamma_5\right)$.

As mentioned in the introduction we will test the method on particularly
complicated situations, including non-self dual configurations and small
instantons.

\begin{figure}
\centering
\includegraphics[width=6 cm,angle=270]{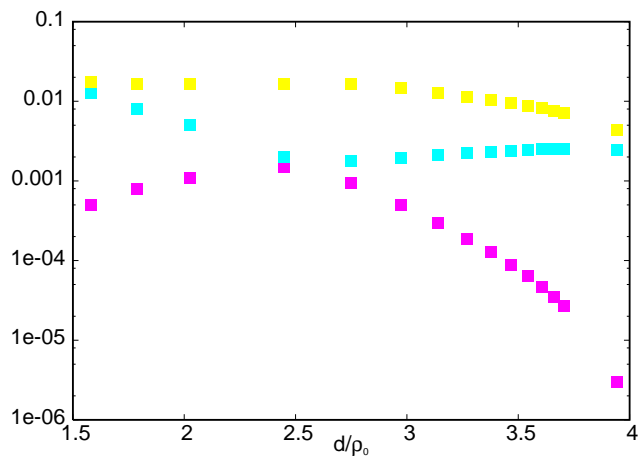}
\caption{Spectrum of lower modes of an IA pair.
Each point represents two positive and two negative
chirality  modes.  For large separation the
super-symmetric modes are the lowest (pink) modes in the spectrum.
This ceases to be the case after the level crossing at
$d \sim 2.5 \rho_0$ ($\rho_0=3a$ is the original instanton size).}
\label{fig:ins_anti_spec}
\end{figure}

\begin{figure}
\centering
\includegraphics[width=8.cm,angle=270]{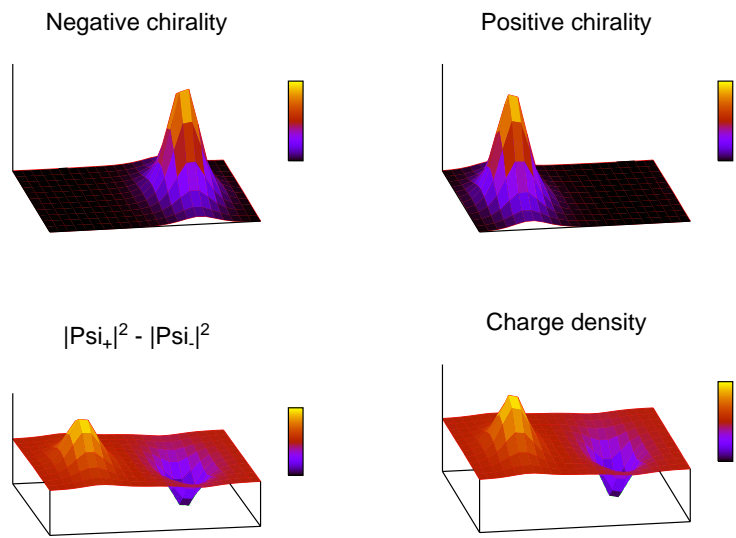}
\caption{Top: Positive and negative chirality lowest modes for a well separated
instanton-anti-instanton pair. Bottom: Reconstruction of the charge density
out of the lowest modes compared with the original one.}
\label{fig:example_ia}
\end{figure}

\section{Testing the method on instanton-anti-instanton pairs.}

Instanton - anti-instanton (IA) pairs are non-self-dual configurations with
zero topological charge. Although such configurations do not have exact zero
modes, we expect to find quasi zero-modes for well separated pairs. In 
particular, we expect 4 positive and 4 negative chirality zero-modes associated
to the isolated instanton and anti-instanton. Indeed, our results for large 
separation, presented in Fig.~\ref{fig:ins_anti_spec}, show a clear gap 
between the lowest eigenvalues and the rest of the spectrum.
The degeneracy of these lowest modes is, however, not correct. We find
only two positive and two negative chirality solutions corresponding,
at large separation, to the super-symmetric zero-modes. 
Figure ~\ref{fig:example_ia} shows the densities of these modes.
Despite the discrepancy in the degeneracy of modes, the 
self-dual and anti-self dual parts of the 
action density can be extracted with good accuracy and the reconstruction 
of the original charge density turns out to be very good.

 The situation becomes less clear as the instanton and anti-instanton
begin to overlap strongly (a sequence of configurations at different 
separations can be generated by improved cooling  ~\cite{epscool}).
In Fig.~\ref{fig:ins_anti_spec}, we can see how the lowest eigenvalues
grow as the separation decreases. There is a critical distance 
($d = 2.5\rho_0$) where the gap in the spectrum closes. In addition to 
this, there is level crossing and the super-symmetric modes cease to be 
the lowest ones in the spectrum. Nevertheless, even in such case,
the super-symmetric modes reconstruct the charge density profile in
good agreement with the original one.
 
\section{Testing the method on rough configurations.}

\begin{figure}
\begin{minipage}[b]{0.48\linewidth} 
\centering
\includegraphics[width=5.cm,angle=270]{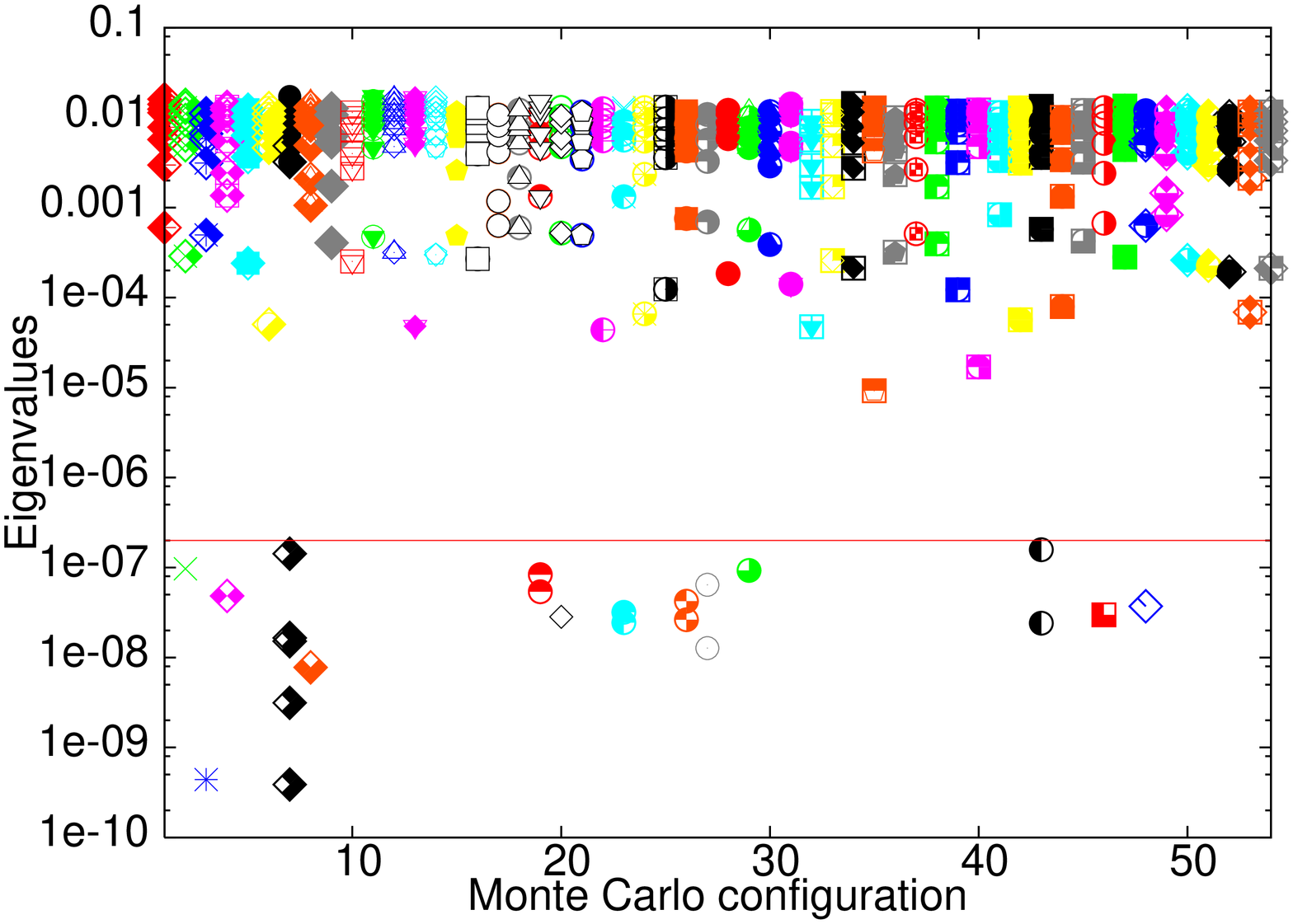}
\caption{Spectra of $D_{ov}(m_{WD}\!=\!1.4)$ for the Monte Carlo
configurations. Points below the straight line are compatible with zero.}
\label{fig:espectro}
\end{minipage}
\hspace{0.4cm}  
\begin{minipage}[b]{0.48\linewidth}
\centering
\includegraphics[width=5.cm,angle=270]{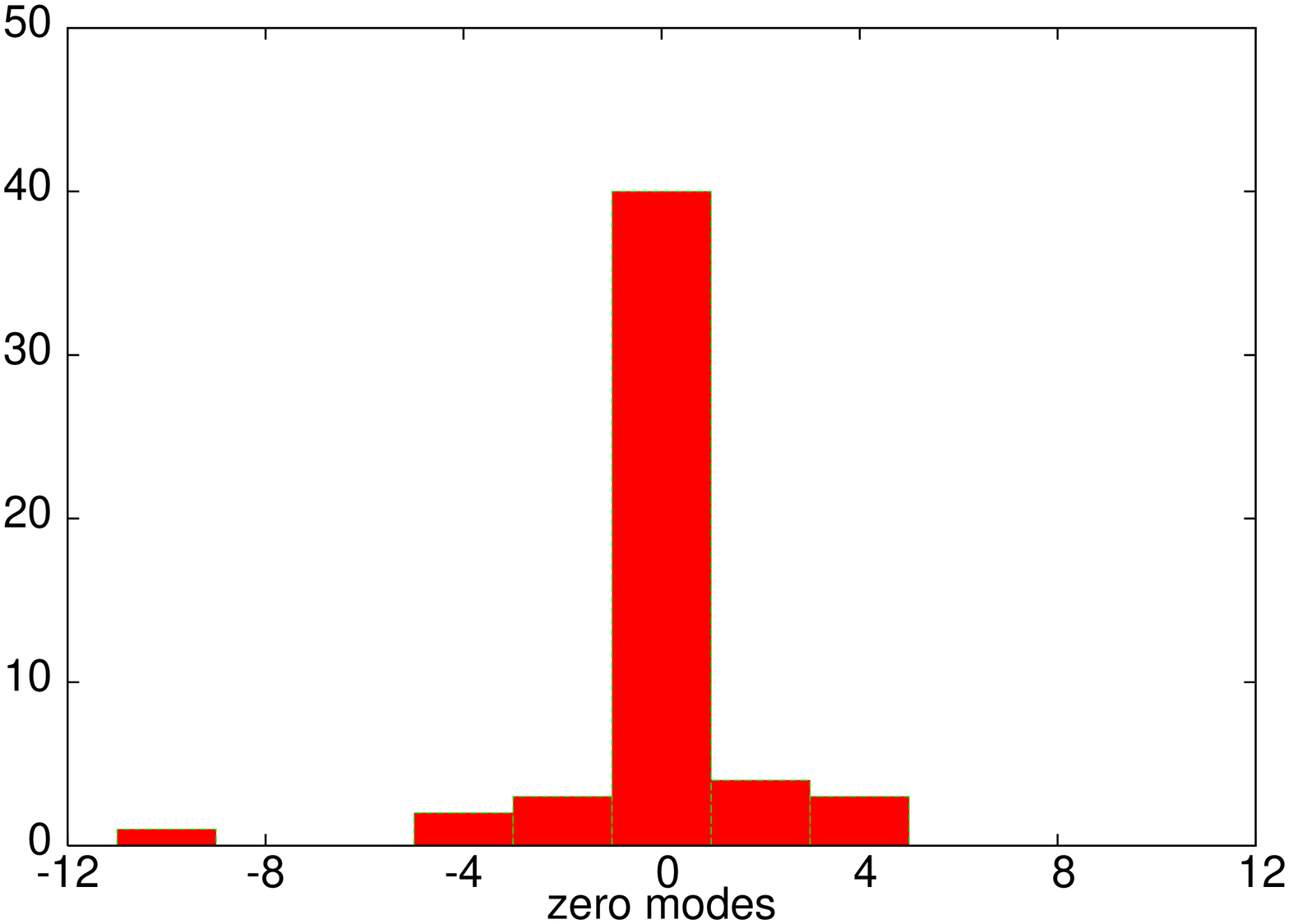}
\caption{Histogram of the number of configurations with a given
number of zero modes. Negative number corresponds to negative 
chirality.}
\label{fig:histo}
\end{minipage}
\end{figure}

As an ongoing part of the project, we have started analyzing
configurations extracted from a SU(2) thermalized ensemble. They 
have been generated  with the Wilson plaquette action, on a $12^4$ 
lattice for $\beta = 2.57$ ($a=0.08$ fm).
For fifty Monte-Carlo configurations we have extracted the 
twelve lowest eigenvalues and eigenvectors. The eigenvalues are 
collected in Fig.~\ref{fig:espectro}, showing a clear gap between
zero and non-zero modes. A histogram of the number of zero
modes is presented in Fig.~\ref{fig:histo}. Surprisingly a significant
fraction of the configurations shows only two zero modes, while the continuum
index theorem predicts $4N, \ N \in Z\!\!\!Z$.
As mentioned before, this mismatch was previously reported by 
Edwards, Heller and Narayanan \cite{EHN}, who interpreted  it
as evidence for fractional topological charge (not expected on the continuum 
for the periodic boundary conditions used).
Contrary to their belief, we will argue that this effect is an artefact 
associated to the roughness of the lattice configurations.
We will  show that in some cases it can be related to the presence 
of small instantons with sizes of the order of the lattice spacing.

\begin{figure}
\begin{minipage}[b]{0.48\linewidth}
\centering
\includegraphics[width=5.cm,angle=270]{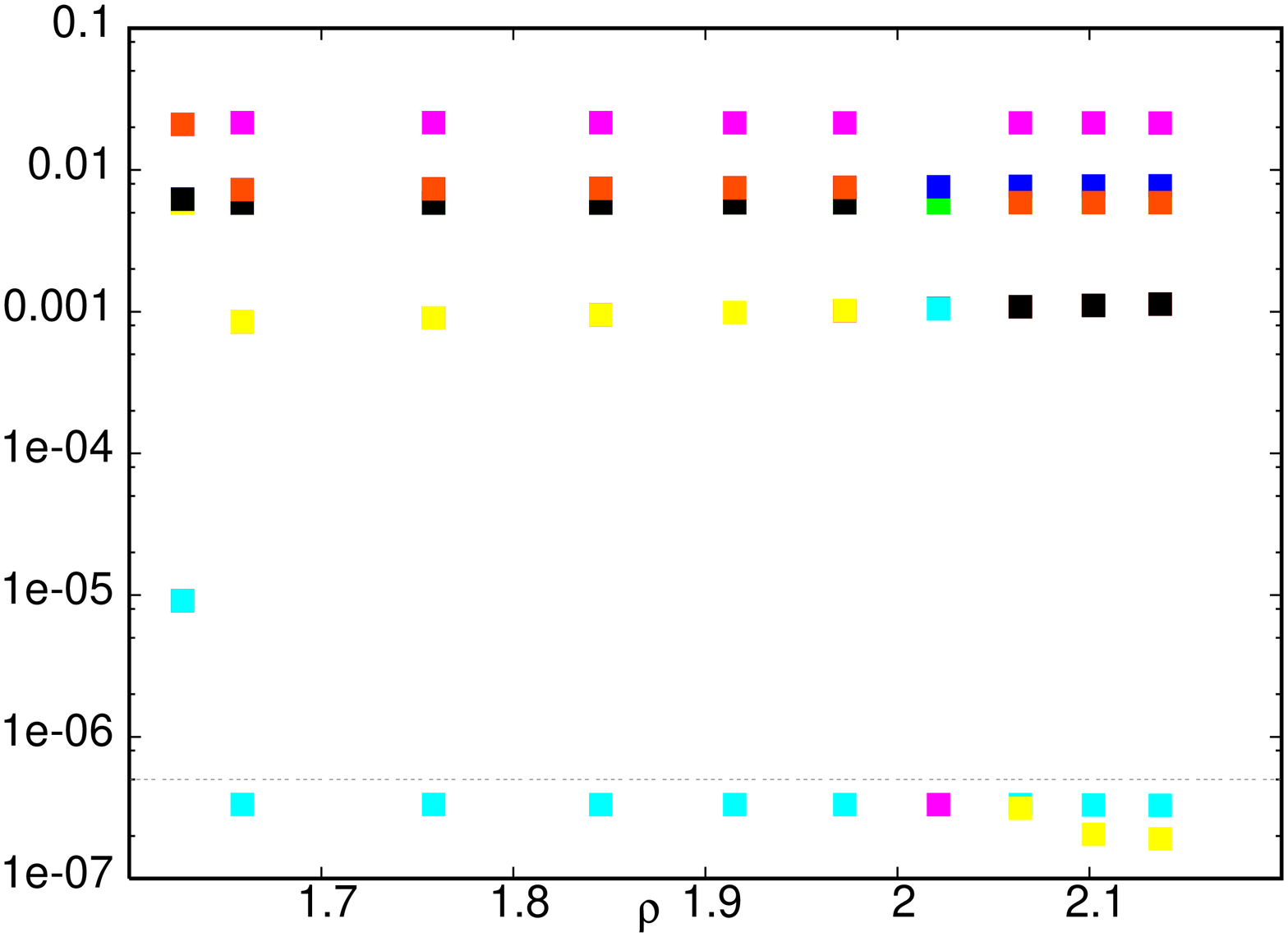}
\caption{Spectrum of  $D_{ov}(m_{WD}\!=\!1.4)$
versus the instanton size in lattice units. Points below 
the line are doubly degenerate and compatible with zero.}
\label{fig:espectro1}
\end{minipage}
\hspace{0.4cm}
\begin{minipage}[b]{0.48\linewidth}
\centering
\includegraphics[width=5cm,angle=270]{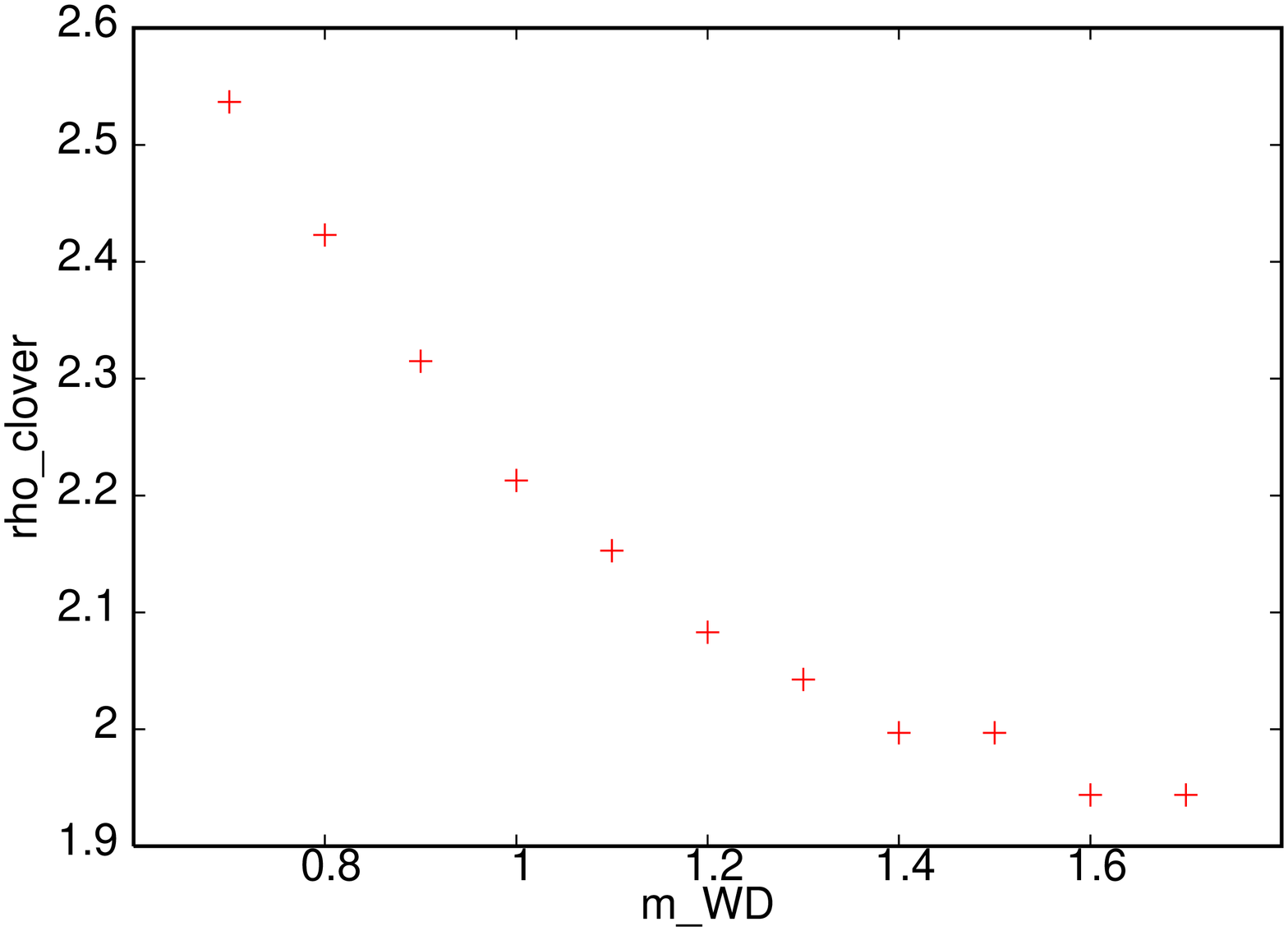}
\caption{Dependence of $\rho_c$ on the mass of the overlap operator.
$\rho_c$ is the critical size (in lattice units) below which the 
spectrum has 2 zero modes.}
\label{fig:mass_rho}
\end{minipage}
\end{figure}

To analyse the behaviour of small instantons we start with a smooth $SU(2)$ 
configuration with $Q=1$. After applying to it several 
cooling sweeps with $\epsilon = 1$ we obtain a sequence of configurations
for different instanton sizes. On them,  we compute the eight first 
adjoint eigenvalues using $m_{WD} = 1.4$ in the overlap 
operator, see Eq. (\ref{eq:overlap}). 
The Atiyah-Singer Index theorem  predicts in this case 
four zero-modes, in two independents pairs $(\psi,\psi_c)$. 
Figure \ref{fig:espectro1} shows the spectrum as a function 
of the  instanton size. Indeed, the initial configuration has 4 zero-modes. 
However, there is a critical size
($\rho_c=2.05 a$) below which only two zero modes remain.
Fig.~\ref{fig:mass_rho} shows the dependence of 
$\rho_c$ on the mass of the overlap operator. 
It  is rather mild for $m_{WD} > 1.4$.
We can use this fact to select an
optimal value of $m_{WD}$  for the analysis of the MC simulations.
The two spurious zero modes dissappear for $\rho < 1.65 a$, quite
independently of the value of $m_{WD}$. Using
$m_{WD} > 1.4$, for which $\rho_c \sim 1.9 a$,
minimises, hence,  the window for spurious eigenvalues. 

\begin{figure}
\begin{minipage}[b]{0.48\linewidth}
\centering
\includegraphics[width=5.cm,angle=270]{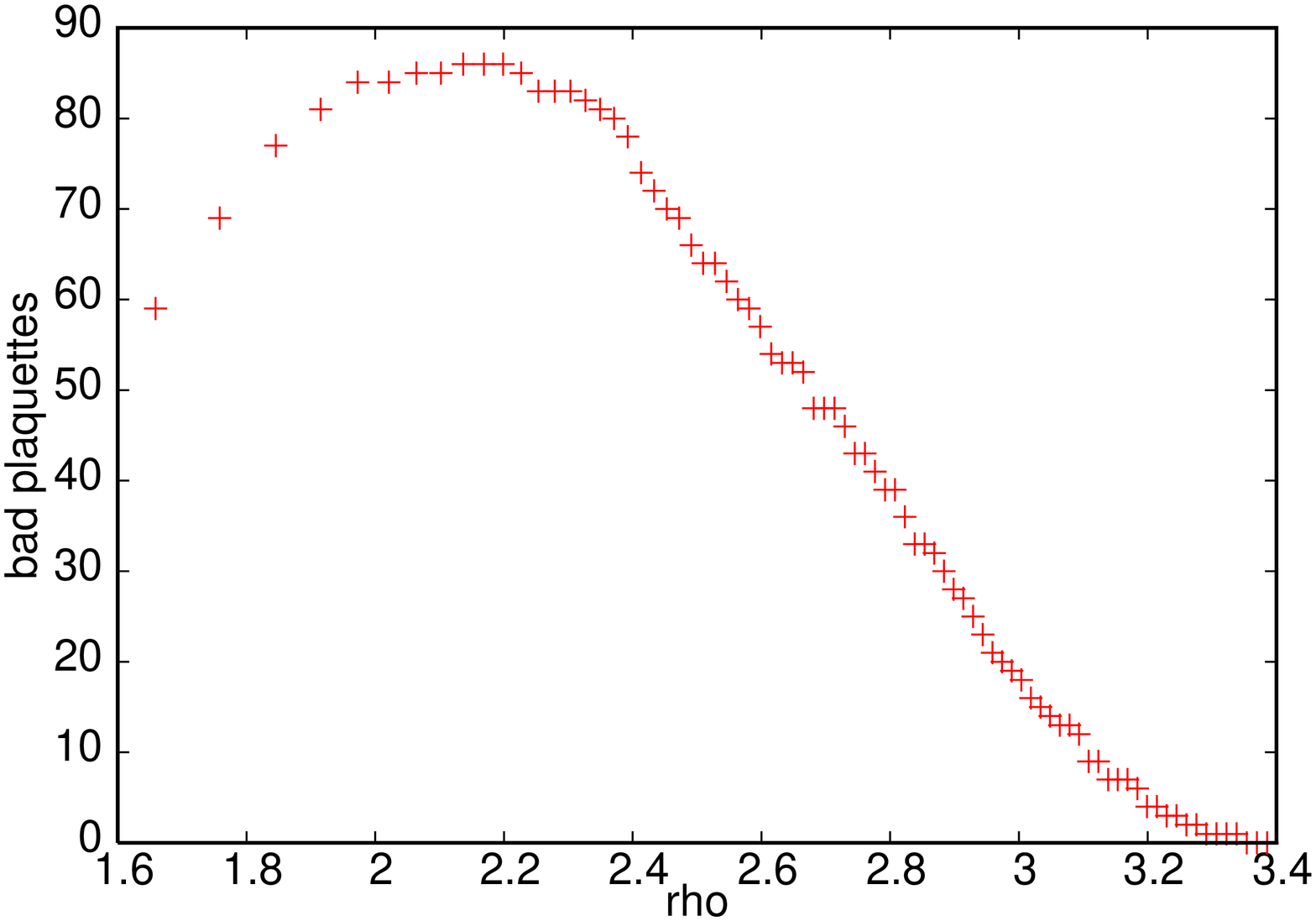}
\caption{Number of plaquettes that violate
the admissibility condition Eq. (4.1) in terms of the
instanton size. $\epsilon(1)=1/30$.}
\label{fig:luscher_condition}
\end{minipage}
\hspace{0.4cm}
\begin{minipage}[b]{0.48\linewidth}
\centering
\includegraphics[width=5.cm,angle=270]{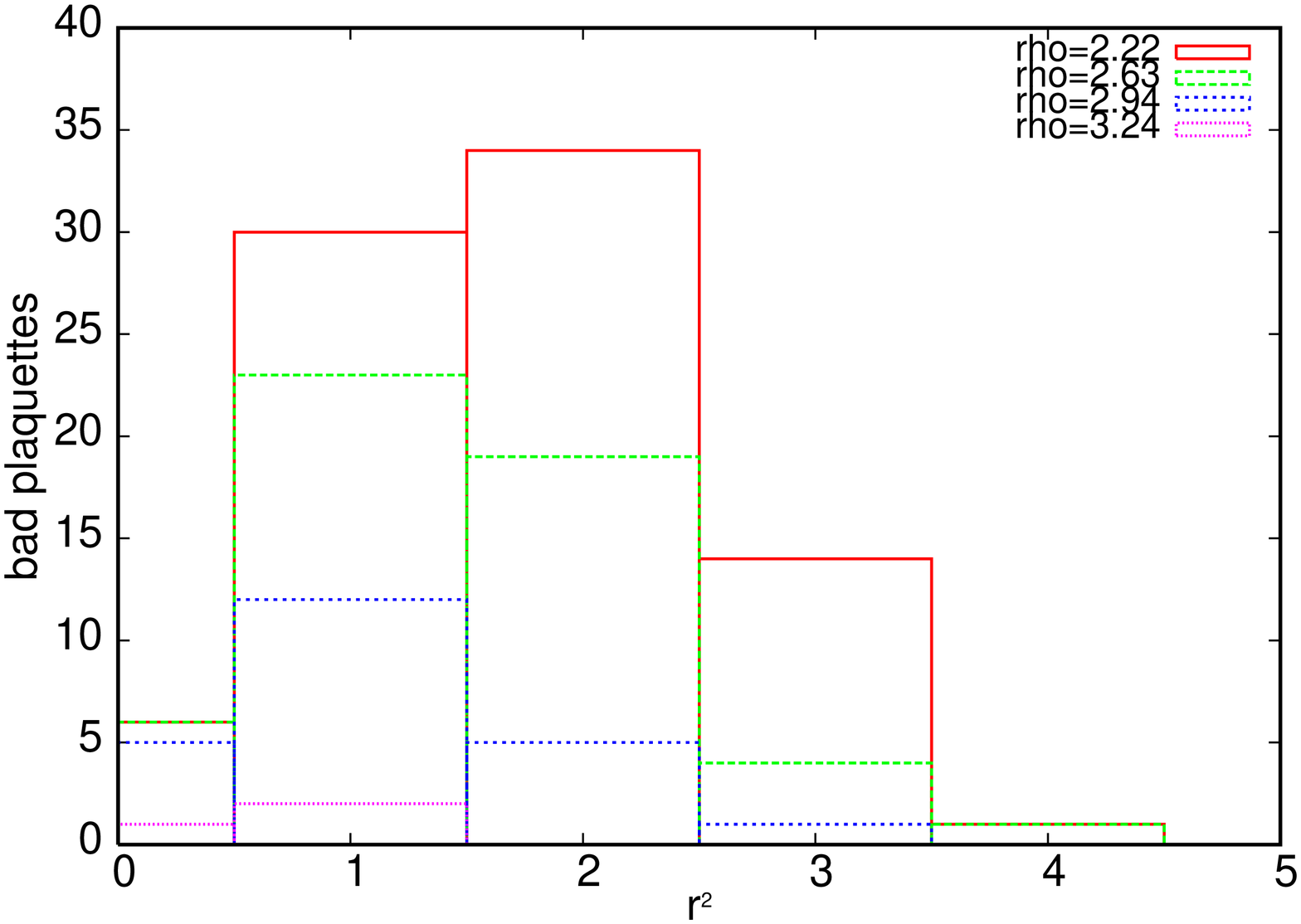}
\caption{Number of plaquettes  that violate the admissibility
condition vs distance from the maximum of the 
action density, for several instanton sizes.}
\label{fig:my_car2}
\end{minipage}
\end{figure}

The spurious zero modes are clearly  associated with
the roughness of the configuration. This can be quantified through
the admissibility condition that should be imposed in order to guarantee 
locality of the overlap operator
\cite{HJL}. This condition, written in terms of the plaquette
$U(p)$, reads
\begin{equation}
\| 1 - U(p) \| \leq \epsilon(m_{WD})\,. \label{eq:luscher_condition}
\end{equation}
Figure ~\ref{fig:luscher_condition} shows the number of plaquettes that violate
Eq. \ref{eq:luscher_condition}, as a function of 
instanton size for the configurations presented in Fig.~\ref{fig:mass_rho}. 
There is a clear link between configurations with spurious zero modes and those 
that violate the admissibility condition. To stress it further we have analysed 
the location of `wrong plaquettes'. Fig.~\ref{fig:my_car2} shows
that they are localized around the  maximum of the action density 
where the configuration becomes rougher for small 
instantons. 

Note that this mismatch is irrelevant if the set of rough configurations
has zero measure in the continuum limit. Our $\beta=2.57$ set has been 
generated with the SU(2) Wilson action. Indeed for this action, general 
arguments, due to Pugh and Teper \cite{pt}, indicate a divergent 
contribution of small instantons in the continuum limit. 
The problem can be avoided by using
an improved action for the Monte-Carlo generation and
by tuning the mass in the overlap operator.
Similar artefacts in the spectrum of the overlap Dirac operator in the 
fundamental representation have been reported in \cite{anna,ilg}. All
this is an indication that the use of improved actions might be essential 
even for the computation of global quantities as the topological susceptibility.

\section{Conclusions.}

We have argued that the adjoint zero-modes of the Dirac 
operator provide an efficient way to extract the topological content
of the Yang-Mills vacuum. The method requires the evaluation of
a small numer of eigenvectors, reducing the computational cost wtih respect
to approaches based on the fundamental representation.
It makes use of the properties of a special set of zero modes which, for 
classical configurations, are directly linked to the gauge action density.
We have probed the goodness of the method in reproducing the charge density
of, non-self-dual, instanton anti-instanton pairs. We have also analysed
the case of rough configurations as small instantons. We have pointed out that
a mismatch between the observed number of zero modes and the continuum
prediction generically arises for instantons below a critical size of the order
of the lattice spacing. 
 
\acknowledgments
We acknowledge financial support from the Madrid Regional Government (CAM), under the 
program HEPHACOS  P-ESP-00346, and  the Spanish Research Ministry (MEC), under contracts \linebreak FPA2006-05807,  FPA2006-05485, FPA2006-05423. Also acknowledged is 
the use of the MareNostrum supercomputer at the BSC-CNS and the  IFT-UAM/CSIC computation cluster.

\end{document}